\documentclass[12pt]{iopart}
\pdfoutput=1
\usepackage[utf8]{inputenc} 			
\usepackage[T1]{fontenc}				
\usepackage{lmodern}					
\usepackage[pdftex]{graphicx} 		    
\usepackage[svgnames]{xcolor}
\usepackage{cite}
\usepackage[pdftex,bookmarks=true, bookmarksnumbered=true,pdfstartview=Fit]{hyperref}
\usepackage{iopams}                     
\usepackage{setstack}                   
\usepackage[english]{babel} 			

\newcommand{\eps}{\varepsilon}							
\newcommand{\dd}{\rmd}									
\newcommand{\ee}{\rme}		        		     		
\newcommand{\moy}[1]{\langle#1\rangle}					
\newcommand{\norm}[1]{\left\|#1\right\|}				
\newcommand{\td}[2]{\frac{\dd #1}{\dd #2}}				
\newcommand{\pd}[2]{\frac{\partial #1}{\partial #2}}	
\newcommand{\eqref}[1]{\eref{#1}}

\newcommand{\revision}[1]{#1}

\begin{document}

\title{Lyapunov exponents of stochastic systems---from micro to macro.}

\author{Tanguy Laffargue, 
		Julien Tailleur and 
		Frédéric van Wijland}

\address{Laboratoire Matière et Systèmes Complexes, 
              UMR 7057 CNRS/P7, 
              Université Paris Diderot, 
              10 rue Alice Domon et Léonie Duquet, 
              75205 Paris Cedex 13, 
              France}

\ead{\mailto{tanguy.laffargue@univ-paris-diderot.fr}}

\begin{abstract}
  Lyapunov exponents of dynamical systems are defined from the rates
  of divergence of nearby trajectories. For stochastic systems, one
  typically assumes that these trajectories are generated under the
  ``same noise realization''. The purpose of this work is to
  critically examine what this expression means. For Brownian
  particles, we consider two natural interpretations of the noise:
  intrinsic to the particles or stemming from the fluctuations of the
  environment. We show how they lead to different distributions of the
  largest Lyapunov exponent as well as different fluctuating
  hydrodynamics for the collective density field. We discuss, both at
  microscopic and macroscopic levels, the limits in which these noise
  prescriptions become equivalent. We close this paper by providing an
  estimate of the largest Lyapunov exponent and of its fluctuations
  for interacting particles evolving with the Dean-Kawasaki dynamics.

\end{abstract}


\maketitle

\section{Introduction}
For a given deterministic dynamical system, Lyapunov exponents tell us
about the rate at which two copies of the system prepared with
close-by initial conditions exponentially diverge from each other in
the course of time. The latter rate needs not be identical along all
phase space directions, hence the existence of a whole spectrum of
exponents.  Extending the notion of Lyapunov exponents to systems
evolving under the action of some external noise has been carried out
long ago~\cite{Benzi1985,Arnold1986, Arnold1988, Grassberger1988,
  Graham1988}. One possible approach, which we henceforth adopt, is to
view the noise as an external perturbation that acts in an identical
way on the infinitesimally close realizations of the system. The
purpose of this work is to critically examine this definition of the
Lyapunov exponents (and in particular when applied to the largest one)
for systems endowed with stochastic dynamics. A number of ambiguities
are still pending, and they boil down to exactly what we mean by
``same realization of the noise''.

Let us now define the largest Lyapunov exponent. We consider a system
described by the a vector $x$ which evolves with the equation
\begin{equation}
	\dot{x} = f(x).
	\label{eq:evo}
\end{equation}
We consider now two copies $x_{A}$ and $x_{B}$ of this system, which
evolves both with the equation \eref{eq:evo}. If $f$ is stochastic,
both copies evolves with the same noise realization. We suppose the
difference between the two copies is small, so the difference $u = x_A
- x_B$ evolves according to
\begin{equation}
	\dot{u} = \mathrm{J}_{f}(x) \,u
\end{equation}
where $\mathrm{J}_f(x)$ is the Jacobian matrix of $f$ in $x$. The
(largest) Lyapunov exponent is defined by
\begin{equation}\label{eqn:deflambda}
	\lambda(t) = \frac{1}{t} \ln \frac{\norm{u(t)}}{\norm{u(0)}}.
\end{equation}
\revision{where $\norm{.}$ stands for the $L^2$-norm.}

With this definition, we will show in section~\ref{sec:LE_1_BP} that
taking ``the same noise realisation'' for the two nearby copies $x_A$
and $x_B$ remains ambiguous, even for simple Brownian particles. We
will consider two different interpretations of the noise entering our
stochastic modelling. In the first case, we will consider a noise
intrinsic to each particle, as was frequently done in the
literature~\cite{Benzi1985}, which we refer to as ``particle-based
noise''. Considering the same noise realization then means that $x_A$
and $x_B$ experience at time $t$ the same noise, say $\eta(t)$, in
both realizations, independently of their positions. Then, we will
introduce an ``environment-based noise'', in which we assume the
origin of the noise to lie solely in the statistics of the surrounding
fluids. Taking the same noise realizations then means that particles
at position $x$ at time $t$ experience the same noise $\chi(x,t)$ in
both realizations. As we show in section~\ref{sec:LE_1_BP}, these two
different interpretations lead to different distributions of
$\lambda(t)$, even for simple Brownian particles\revision{, whereas it
  is impossible to discriminate between these two prescriptions when
  looking at a single copy of the system}.

We then turn to the study of the collective dynamics of Brownian
particles in section~\ref{sec:colldyn}. Following the approach of
Dean~\cite{Dean1996}, we construct the fluctuating hydrodynamics for
the density fields corresponding to the two types of
noise. Interestingly, these hydrodynamics are typically different,
except in the limit where the spatial correlation length of the
environment-based noise is much smaller than the interparticle
distance.

Next, we define the largest Lyapunov exponent associated to the
collective density field in section~\ref{sec:tdcd}. To do so, we first
verify that linearizing the fluctuating hydrodynamics is equivalent to
coarse-graing directly the microscopic tangent dynamics. Again, for
the two types of noise to be equivalent, the spatial correlation
length of the environment-based noise needs to be much smaller than
the interparticle distance. This is, however, not sufficient and one
also needs to compare initial conditions that are separated by a
distance larger than the noise correlation length when constructing
the tangent dynamics.

Sections~\ref{sec:LE_1_BP} to~\ref{sec:tdcd} thus allow us to
unambiguously define the Lyapunov exponent associated to the
collective density field of interacting particles. We then provide in
section~\ref{sec:interactions} \revision{two estimates of its mean
  values, one at a purely mean-field level and the other by retaining
  the Gaussian fluctuations of the density field}.

\section{Lyapunov exponent of a Brownian particle \label{sec:LE_1_BP}}

A single particle undergoing Brownian diffusion in a solvent is
arguably the simplest of stochastic systems. In this section, we show
that the definition of its Lyapunov exponent, even once the ``same
noise convention'' has been taken, is ambiguous. For sake of
simplicity, we consider a one-dimensional system.

\subsection{Noise on the particle \label{sec:LE_1_part_noise_part}}
The standard description of Brownian motion is to consider the
following stochastic differential equation for the position $r(t)$ of
a particle:
\begin{equation}
	\dot{r} = \eta(t)
	\label{eq:bruit particules}
\end{equation}
where $\eta(t)$ is a zero-mean Gaussian white noise whose correlations
satisfy $\moy{\eta(t) \,\eta(t')} = 2 D \,\delta(t-t')$.

To compute the Lyapunov exponent, we consider two copies of our system
evolving with the same noise realization. The tangent vector evolves
according to the equation:
\begin{equation}
  \dot{u} = 0.
\end{equation}
Since the two particles experience the same noise realization, and
since there is nothing else in the system to make their dynamics
differ, the distance between them remains trivially constant and
$\lambda = 0$.

\subsection{Noise on the environment \label{sec:LE_1_part_noise_env}}
\subsubsection{Particle in a Gaussian random field}
For a colloid in a fluid, the origin of the noise lies in the
collision with the fluid particles. It is thus natural to consider a
(Gaussian) noise \textit{field} $\chi(r,t)$ experienced by a particle at position
$r$ at time $t$. The single-particle dynamics then read
\begin{equation} 
  \dot{r} = \chi(r(t), t)
  \label{eq:bruit sites}
\end{equation}
where $\moy{\chi(x,t)}=0$ and $\moy{\chi(x, \,t) \, \chi(x', \,t')} =
2 D \,C(x-x') \,\delta(t-t')$. The function $C(x)$ represents the
spatial correlations of the fluctuations in the fluid. For simplicity,
we choose it to be smooth and even (to respect isotropy), which implies
$C'(0) = 0$. 

In addition, we expect the correlation function $C(x)$ to be maximal
at $x=0$ and to decrease as $\left| x \right|$ increases. This
implies ${C''(0) \leqslant 0}$, which we use to define a characteristic length scale $\ell$ by  ${C''(0) \equiv
  -{1}/{\ell^{2}}}$. Furthermore, we normalize $C(0)=1$ so that the noise
amplitude is solely controlled by $D$. \revision{At this point, whenever the function $C$ respects the previous constraints, it is impossible to say if the noise is particle-- or environment--based.}

Equation~\eref{eq:bruit sites} can be rewritten as
\begin{equation}
  \label{eqn:SDE1}
	\dot{r} = \int \dd y\, \delta(y-r(t)) \, \chi(y, \,t),
\end{equation}
Equation~\eref{eqn:SDE1} involves a multiplicative noise, which in principle calls for a specification of the time-discretization we resort to. As we show in~\ref{app:NTD}, a simplifying mathematical feature makes the discretization an irrelevant feature as long as $C'(0)=0$. We use the Stratonovich
convention in this section, so that the standard rules of differential
calculus apply.

Let us now compare equations \eref{eq:bruit particules} and
\eref{eq:bruit sites}. Both
equations \eref{eq:bruit particules} and \eref{eqn:SDE1} have their first Kramers-Moyal coefficient equal to zero
(see~\ref{app:NTD}) while their second coefficient expectedly coincide:
\begin{equation}
  \lim_{\Delta t \to 0} \frac{\moy{[r(t+\Delta t)-r(t)]^2}}{\Delta t} = 2 D
\end{equation}
Higher order Kramers-Moyal coefficients scale with higher orders of
$\Delta t$ and hence both dynamics lead to the same Fokker-Planck. One
could thus naively expect their Lyapunov exponents to be equal. This
is what we investigate in the next subsection.

\subsubsection{Calculation of the Lyapunov exponent.}
We now consider two infinitesimally close initial conditions, $r_1(0)$ et
$r_2(0)$, which evolve with the same noise realization
$\chi(y,t)$. The evolution of $u(t) = r_1(t) - r_2(t)$ is given by the
linearized (tangent) dynamics
\begin{equation}
	\dot{u} = - u(t) \int \dd y \, \chi(y, \,t) \, \partial_{y}\delta(y-r(t)). 
\end{equation}
whose solution reads
\begin{equation}
  u(t)=u(0) \exp[- \int_0^t \dd t' \int \dd y \, \chi(y, \,t') \, \partial_{y}\delta(y-r(t'))]. 
\end{equation}
The Lyapunov exponent is thus given by
\begin{equation}
  \lambda(t) = - \frac{1}{t} \int_{0}^{t}\dd t' \int \dd y \, \chi(y, \,t') \, \partial_{y}\delta(y-r(t')) .
\end{equation}
This is a fluctuating quantity. The global expansion coefficient
$\Lambda(t) = t \lambda(t)$ actually satisfies another Langevin equation
\begin{equation}
  \dot \Lambda(t) = - \int \dd y \, \chi(y, \,t) \, \partial_{y}\delta(y-r(t)).
\end{equation}
To compute the probability distribution of $\lambda(t)$, we will thus
compute and solve the Fokker-Planck equation satisfied by the joint probability
$P(r,\Lambda,t)$. Let us compute the first and second order coefficients
of the Kramers-Moyal expansion of the coupled variables $\Lambda$ and $r$. As we have chosen
a Stratonovich convention, we have
\begin{eqnarray}
	\Delta r &= \int_{t}^{t + \Delta t}\dd t' \int \dd y \, \delta(y - r(t) - \frac{1}{2} \Delta r)) \, \chi(y, \,t')\\
	\Delta \Lambda &= - \int_{t}^{t + \Delta t}\dd t' \int \dd y \, \partial_y \delta(y - r(t) - \frac{1}{2} \Delta r)) \,  \chi(y, \,t')
\end{eqnarray}
which allows us to compute
\begin{eqnarray}
	&\lim_{\Delta t \to 0} \frac{\moy{\Delta \Lambda}}{\Delta t}   = D C''(0) = - \frac{D}{\ell^{2}}\\
	&\lim_{\Delta t \to 0} \frac{\moy{\Delta \Lambda^2}}{\Delta t} = - 2  D C''(0) = 2 \frac{D}{\ell^{2}} \label{eq:variance lambda}\\
	&\lim_{\Delta t \to 0} \frac{\moy{\Delta \Lambda \, \Delta r}}{\Delta t} = 2 D C'(0) = 0
\end{eqnarray}
The correlation between $\Delta \Lambda$ and $\Delta r$ is zero here,
so we actually need not consider the joint probability $P(r,\,
\Lambda,\,t)$ and we can establish a Fokker-Planck equation for
$P(\Lambda,\,t)$ only:
\begin{equation}
	\partial_{t} P(\Lambda,\,t) = \frac{D}{\ell^{2}} \, (\partial_{\Lambda} + \partial_{\Lambda}^2) \,P(\Lambda,\,t).
	\label{eq:P lambda}
\end{equation}
We recognize the Fokker-Planck equation of a Brownian particle with 
position $\Lambda$, starting from the origin $\Lambda(0)=0$ and subject to a constant force, which simply diffuses in the co-moving
frame with velocity $D/\ell^2$, and with diffusion constant $D/\ell^2$. We thus have that 
\begin{equation}
  P(\Lambda,t)=\frac{1}{\sqrt{4\pi D\ell^{-2} t}}\ee^{-\frac{(\Lambda-D\ell^{-2}t)^2}{4 D\ell^{-2} t}}
	\label{eq:fonction gene Lambda}
\end{equation}
By changing variable from $\Lambda$ to $\lambda = \Lambda/t$ we can directly read off the large deviation function of $\Lambda=\lambda t$ as the rate of increase of $P(\lambda,t)$:
\begin{equation}
  P(\lambda, \, t) = \sqrt{\frac{t \ell^{2}}{4 \pi D}} \exp\Big[{- \frac{t \ell^{2}}{4 D} \left( \lambda + \frac{D}{\ell^{2}} \right)^{2}}\,\Big].\label{eqn:PofLenv}
\end{equation}
The Lyapunov exponent $\lambda(t)$ is thus non zero, not even on
average, given that $\moy{\lambda} = - {D}/{\ell^{2}} < 0$. Decreasing
the spatial correlations of the noise leads two nearby particles to
follow different subsequent trajectories, and it is thus expected that
taking $\ell$ finite allows for fluctuations of $\lambda$. It was,
however, more difficult to predict that decreasing $\ell$ was going to
increase the \textit{mean} Lyapunov exponent and make the system more
stable on average: the noise could have make both particles pull away
from each other.

Note that the probability distribution~\eref{eqn:PofLenv} depends on
the spatial correlations of the noise solely through its short-scale
structure: it does not depends on the full spatial correlation
function but only on $\ell^2=-1/C''(0)$. In a single particle problem,
having a noise ``attached'' to a particle, as in equation~\eref{eq:bruit
  particules}, is equivalent to applying the same noise at every location
in space, i.e.  to $C(x) = 1$ for all $x$. This in turn implies
$\ell = + \infty$ and our two results are then consistent with each other since
\begin{equation}
	P(\lambda, \, t) \underset{\ell \to +\infty}{\sim} \delta(\lambda).
\end{equation} 

Changing the nature of the noise thus completely modifies the tangent
dynamics, even though the two individual dynamics~\eref{eq:bruit particules} and
\eref{eq:bruit sites} yield the same Fokker-Planck equation. A
prescription which appeared purely philosophical for the dynamics of a
single copy of the system turns out to have important consequences
when looking at the Lyapunov dynamical stability of the system.

\section{Collective dynamics of Brownian particles}
\label{sec:colldyn}

Let us now consider $N$ identical particles which we endow with either
of the noise prescriptions discussed in the previous section. The
local fluctuating density field, defined by $\rho(x,t)=\sum_i
\delta(x-r_i(t))$, where $r_i(t)$ denotes the position of particle $i$
at time~$t$, evolves according to some stochastic dynamics.

For particle-based individual noise, this is the well-known
Dean-Kawasaki Langevin equation~\cite{Dean1996} that governs the
evolution of $\rho$, which we briefly rederive in
section~\ref{sec:noiseindpart}. We then construct the fluctuating
hydrodynamics stemming from an environment-based noise in
section~\ref{sec:noiseenvpart}. Last, we compare the two dynamics in
section~\ref{sec:comparison}.

As we have just shown, the individual dynamics leads to distinctly
different tangent dynamics for the two types of noise, except when the
correlation length $\ell$ of the environment-based noise is
infinite. It may thus be somewhat of a surprise that the dynamics of
the collective local density field become equivalent in the converse
limit, when the correlation length $\ell$ is smaller than the
interparticle distance~$d$. It is indeed a nontrivial result that, in
the limit $\ell \ll d$, the Dean-Kawasaki equation also governs the
evolution of $\rho$ when individual dyamics are driven by a noise field
as in \eref{eqn:SDE1}. 

In the general case, when $d$ and $\ell$ are comparable, the
fluctuations of the density field scale differently for the two types
of noise. For particle-based noise, the fluctuations of each particle
sum up incoherently, and one recovers a noise variance proportional to
$\rho(x)$. On the contrary, for environment-based noise, the
fluctuations become correlated and the noise variance is thus
proportional to $\rho^2(x)$. \revision{Our goal here is not to study the general case of environment-based noises with long-range correlations~\cite{Nardini2012, Nardini2012a, Bouchet2013}. We want two nearby particles to experience independent noises, but we want two particles at the same position in two copies of our system to experience the same noise. This is why we consider a noise field with $\ell \ll d$.}

\subsection{Noise on the particles}
\label{sec:noiseindpart}
Consider $N$ Brownian particles such that particle $j$ is subjected to a
zero-mean Gaussian white noise $\eta_j$ and evolves according
to
\begin{equation}
  \dot r_j(t) = \eta_j(t).
\end{equation}
The particles are uncorrelated so that $\moy{\eta_i(t) \,\eta_j(t')} =
2 D \,\delta_{i, j}\, \delta(t-t')$. We want to determine how the
collective mode associated to the density of particles
\begin{equation}
	\rho(x, t) = \sum_{j=1}^{N} \rho_{j}(x, t) \qquad $with$ \qquad  \rho_{j}(x,t) = \delta(x-r_{j}(t))
\end{equation}
evolves. This was done by Dean using It\=o calculus~\cite{Dean1996}
and a method that we will follow in the next section. We use a
slightly less rigorous path here, which is however more easily
generalized. For sake of completeness, we consider the case of a
multiplicative noise by allowing $D$ to depend on $r_j$. We treat
$\rho$ as a multi-dimensional function of the $r_j(t)$'s and use It\=o
lemma to compute its time-evolution~\footnote{This is the non-rigorous
  part since It\=o lemma applies for twice-differentiable scalar
  function, which the $\rho_j$'s are not.}
\begin{equation}
  \dot \rho(x,r_j(t)) = \sum_{j=1}^{N}   \dot r_j\,\partial_{r_j} \rho + \sum_{j=1}^{N}   D(r_j)\,\partial_{r_j}^2 \rho.
\end{equation}
Using that $\partial_{r_j}\rho=\partial_{r_j}\rho_j$, one gets
\begin{equation}
  \dot \rho(x,r_j(t)) = \sum_{j=1}^{N}   \partial_{r_j} \delta (x-r_j) \eta_j + \sum_{j=1}^{N}   D(r_j) \partial_{r_j}^2 \delta(x-r_j).
\end{equation}
We then use that $\partial_{r_j}\delta(x-r_j)=-\partial_x \delta(x-r_j)$ to get
\begin{equation}
  \dot \rho(x,r_j(t)) = \partial_x \Big[ -\sum_{j=1}^{N}  \delta (x-r_j) \eta_j + \partial_x \sum_{j=1}^{N}   D(r_j) \delta(x-r_j) \Big].
\end{equation}
Finally, using that $\sum_{j=1}^{N}  D(r_j) \,\delta(x-r_j) = \sum_{j=1}^{N}  D(x) \,
\delta(x-r_j) = D(x) \, \rho(x)$, we find the generalization of Dean's
results to multiplicative noise
\begin{equation}
 \partial_t \rho(x,t) = \partial_{x}^{2}[D(x) \rho(x,t)] - \partial_x \xi(x,t) \quad $with$ \quad \xi =  \displaystyle\sum_{j=1}^{N} \rho_j(x, t) \,\eta_j(t)\label{eqn:pasbelle}
\end{equation}
The noise $\xi(x,t)$ is a sum of Gaussian noises and hence Gaussian. It is
completely characterized by its average, which is 0, and its variance,
which is
\begin{eqnarray}
  \moy{\xi(x, t) \, \xi(x', t')} &=  \sum_{i,j=1}^{N}  \rho_{i}(x) \, \rho_{j}(x') \, \moy{\eta_i(t) \, \eta_{j}(t')}\\
								&= 2 \sum_{j=1}^{N} D(r_j) \, \delta(x-r_j(t)) \, \delta(x'-r_{j}(t)) \, \delta(t-t')\\
								&= 2 D(x) \, \delta(x-x') \, \delta(t-t') \sum_{j=1}^{N}  \delta(x'-r_{j}(t)) \\
								&= 2  D(x)\,\rho(x, t) \,\delta(x-x') \, \delta(t-t')
\end{eqnarray}
The evolution of $\rho$ is then self-consistently given by
\begin{equation}
 \partial_t \rho(x,t) = \partial_{x}^{2}[ D(x)\rho(x,t)] -\partial_x \xi^D(x,t)\;$with$\; \xi^D(x,t) = \sqrt{\rho(x,t)}\, \eta(x,t)
 \label{eq:Dean noise particles}
\end{equation}
where 
\begin{equation}
\moy{\eta(x, t)} =0\quad $and$\quad  \moy{\eta(x, t)\, \eta(x', t')}    =  2  D(x)\, \delta(x-x') \,\delta(t-t').
\end{equation}
From now on, we turn back to the case of constant $D$. Let us now consider
the case of environment-based noise.

\subsection{Noise on the environment}
\label{sec:noiseenvpart}
Consider now the noise field from
section~\ref{sec:LE_1_part_noise_env}, resulting from a fluctuating
environment. To derive the stochastic dynamics of $\rho(x,t)$, we 
again make use the It\=o lemma, now generalized to a random field. For
completeness, an equivalent derivation using Stratonovich convention
is presented in~\ref{app:strato}. Let us consider $N$ particles whose
individual dynamics are
\begin{equation}
	\td{r_{j}}{t} = \chi(r_{j}(t), t)
\end{equation}
where $\chi(x, t)$ is the Gaussian random field defined in
\ref{sec:LE_1_part_noise_env}. As shown in~\ref{app:NTD}, this
equation does not depend of the chosen discretization (It\=o or
Stratonovich) and we use in this section It\=o calculus. As before, we
define individual and global densities as
\begin{eqnarray}
	\rho_{i}(x,t) &= \delta(x - r_{i}(t))\\
	\rho(x, t) &= \sum_{i} \rho_{i}(x,t)
\end{eqnarray}
We could, once again, apply It\=o Lemma directly to $\rho$, but we
follow for completeness the path of Dean~\cite{Dean1996} here.  Let
$f(r_i)$ be an arbitrary function of $r_{i}$. The generalization of
the It\=o lemma to a field of noise tells us that
\begin{equation}
	\td{f(r_i(t))}{t} = f'(r_{i}(t)) \td{r_i}{t} + D \,f''(r_{i}(t))
\end{equation}
where we have used that $C(0)=1$. Using the definition of $\rho_i$,
this can be rewritten as
\begin{eqnarray}
	\td{f(r_i(t))}{t} &= \int \dd x ~\rho_{i}(x, t) \left[ f'(x) \chi(x, t) + D \,f''(x) \right] \\
	&= \int \dd x ~f(x) \left[ -\partial_{x} \left( \rho_{i}(x, t) ~\chi(x, t) \right) + D ~\partial_{x}^{2} \rho_{i} \right]\label{eqn:bla}
\end{eqnarray}	
where the second equalities comes from an integration by part. Since
\begin{equation}
f(r_{i}(t)) = \int \dd x ~\rho_{i}(x, t) f(x)
\end{equation}
one also has that
\begin{equation}
  \td{f(r_i(t))}{t} = \int \dd x ~f(x) ~\partial_{t} \rho_{i}. \label{eqn:bli}
\end{equation}
Since equations~\eref{eqn:bla} and~\eref{eqn:bli} hold for any
function $f$, they yield
\begin{equation}
  \partial_{t} \rho_{i}(x,t) =D ~\partial_{x}^{2} \rho_{i}(x,t) - \partial_{x} \left[ \rho_{i}(x, t) ~\chi(x, t) \right].
\end{equation}
Summing over i and using the definition of the density we obtain
\begin{equation}
	\partial_{t} \rho(x,t) = D ~\partial_{x}^{2} \rho(x,t) -\partial_x \Gamma(x,t)\quad$with$\quad\Gamma(x,t)= \rho(x, t) ~\chi(x, t).
	\label{eq:Dean noise environment}
\end{equation}
An interesting feature of this calculation is that the noise acting on $\rho$ is directly
given as a functional of $\rho$, without any prior reference to the
individual microscopic densities $\delta(x-r_j)$ (as opposed to
equation~\eref{eqn:pasbelle}).

\subsection{Comparison between particle-based and environment-based collective dynamics}
\label{sec:comparison}
Without further physical requirements, the two breeds of noise we have introduced
 do not lead to the same physics for the collective modes, in the same line as what we have already commented regarding the particles's Lyapunov exponents. For the density field, this can be seen by inspecting the noise variance in equations~\eref{eq:Dean noise particles} and~\eref{eq:Dean
 noise environment}.

When each particle has its own noise, the variance of the collective
noise $\xi$ is proportional to $\rho$. This simply comes from the fact
that two nearby particles remain independent. Therefore, their
contributions to the density fluctuations add incoherently. When
particles experience a Gaussian field $\chi$ with finite correlation
length, nearby particles are no longer independent: density
fluctuations are due to the environment and their amplitude scales
with the local density $\rho$. The noise variance is thus proportional
to $\rho^2$.

Note that taking the limit $\ell\to 0$ in~\eref{eq:Dean noise
  environment}, using $C(x) = \delta(x)/\delta(0)$\footnote{The $\delta(0)$
  is present for normalization purpose.}, leads to
\begin{equation}
	\moy{\Gamma(x, t) \, \Gamma(x', t')} = \frac{2D}{\delta(0)} \, \rho(x, t)^2 \, \delta(x-x') \, \delta(t-t') 
\end{equation}
This does not suffice to recover the noise of independent
particles since two point-like particles at the same position $x$ still
experience the same noise.

To properly recover the result of the noise on particles, we have to
introduce the typical size of the particles $d$.  When it is larger
than the typical correlation length of the noise $\ell$, we expect
 to recover independent contribution from each particle to the
fluctuations of the local density field. Since two particles cannot be
correlated, because $\norm{r_{i} - r_{j}} > d \gg \ell$, the
spatial correlation function reduces to
\begin{equation}
  C(r_{i}(t) - r_{j}(t')) = \delta_{i,j} \, C(0) = \delta_{i,j}.
\end{equation}
and the correlations of the noise is then given by
\begin{eqnarray}
	\moy{\Gamma(x, t) \, \Gamma(x', t')} &= 2D \, \delta(t-t')   \sum_{i,j=1}^{N} \delta(x-r_i) \delta(x'-r_j) C(r_j-r_i)\nonumber\\
	&=2D \, \delta(t-t')  \sum_{i,j=1}^{N}  \delta(x-r_i) \delta(x'-r_j) \delta_{i,j}\nonumber\\
	&=2D \, \delta(t-t')   \sum_{i=1}^{N}  \delta(x-r_i) \delta(x'-x)\nonumber\\
	&=2D \,  \rho(x, t) \, \delta(x'-x)\, \delta(t-t')
\end{eqnarray}
which is what we obtained with the noise on particles. 

To summarize, particle-based and environment-based noises yield two
different collective dynamics, mostly because in the latter case
nearby particles are effectively correlated. The proper way to recover
uncorrelated noise for the environment-based noise is thus to
consider particles with finite-size $d$ and take $d \gg \ell$. Note
that this simply amounts to considering cases where $\norm{r_i-r_j}\gg
\ell$, i.e. systems which are dilute at the scale of the fluid's
correlation length.

\section{Tangent dynamics of the collective density mode}
\label{sec:tdcd}
As for the single particle case, we thus expect that the type of noise has an
impact on the divergence of nearby trajectories at the macroscopic scale,
and hence the Lyapunov exponents associated to the collective density
mode $\rho$. Note that it is also unclear whether the tangent dynamics
associated to $\rho$, obtained by linearizing the stochastic partial
differential equation obeyed by $\rho(x,t)$, should coincide with the
direct coarse-graining of microscopic tangent dynamics. In this
section we thus show that linearizing and projecting on the collective
density modes are indeed commuting operations. The approach followed
analytically in~\cite{Laffargue2015} to compute the fluctuations of
the Lyapunov exponent of spatially extended systems incidentally sits on firmer grounds. 

\subsection{Linearizing the fluctuating hydrodynamics}
Let us consider two infinitesimally close initial density profiles $\rho_1(x,0)$
and $\rho_2(x,0)$. The evolution of their difference
$u(x,t)=\rho_1(x,t)-\rho_2(x,t)$ is obtained by linearizing the
equations~\eref{eq:Dean noise particles} and~\eref{eq:Dean noise
  environment}.
\paragraph{Noise on the particles.}
The tangent evolution associated to \eref{eq:Dean noise particles} is
\begin{equation}
	\partial_{t} u(x,t) = D ~\partial_{x}^{2} u(x,t) - \partial_{x} \xi_u^D(x,t)\;$with$\; \xi_u^D(x,t)=\displaystyle\frac{u(x,t)}{2 \sqrt{\rho(x,t)}} ~\eta(x, t)\label{eqn:ulinD}
\end{equation}
where $\eta(x,t)$ is the zero-mean Gaussian white noise defined
in~\eref{eq:Dean noise particles}.

\paragraph{Noise on the environment.}
The tangent evolution associated to \eref{eq:Dean noise environment} is
\begin{equation}\label{eqn:FHUNE}
	\partial_{t} u(x,t)= D ~\partial_{x}^{2} u(x,t)- \partial_{x} \Gamma_u^D(x,t)\;$with$\; \Gamma_u^D(x,t)= u(x, t) ~\chi(x, t)
\end{equation}
and $\chi(x,t)$ is the zero-mean Gaussian white noise defined
in~\eqref{eq:bruit sites}.
\subsection{Starting from microscopic dynamics}
Let us now start from the microscopic dynamics and derive the
hydrodynamic behavior of the differences between two copies of the
system. We take two sets $\{r_j\}$ and $\{r_j + \delta r_j\}$ of
infinitely close initial positions evolving with the same equation of
evolution. We can then define two densities:
\begin{eqnarray}\label{eqn:defdeltar}
	\rho(x, t) &= \sum_{j=1}^{N} \delta(x-r_{j}(t))\\
	\tilde\rho(x, t) &= \sum_{j=1}^{N} \delta(x-r_{j}(t) - \delta r_j(t))
\end{eqnarray}
We want to establish the evolution equation of the (small) differences
between the two density fields
\begin{equation}\label{eqn:defu}
	u(x, t) = \tilde\rho(x, t) - \rho(x, t) \simeq - \sum_{j=1}^{N} \delta r_j(t) \, \partial_{x} \, \delta(x - r_j(t)).
\end{equation}

\subsubsection{Noise on the particles}
For the noise on particles, the positions and perturbations evolve according to
\begin{equation}
  \td{\delta r_j}{t} = 0
\end{equation}
which means that the $\delta r_j$ are constant. Following the path of
section~\ref{sec:noiseindpart}, we now apply the It\=o lemma to u:
\begin{eqnarray}
	\pd{u}{t} 	&= \sum_{j=1}^{N} \pd{u}{r_j} \eta_j + D \sum_{i=1}^{N} \sum_{j=1}^{N} \frac{\partial^2 u}{\partial r_i \, \partial r_j} \delta_{i,j}\\
        &= \sum_{j=1}^{N} \partial_{x}^2 \, \delta r_j \, \delta(x-r_j) \eta_j + D \,\partial_x^2 u
\end{eqnarray}
which can be rewritten as
\begin{equation}\label{eqn:toto12}
	\partial_t u = D \, \partial_x^2 u +\partial_x \xi_{u}(x, t)\quad$with$\quad \xi_u= \partial_{x}\displaystyle\sum_{j=1}^{N} \, \delta r_j \, \delta(x-r_j) \eta_j
\end{equation}
and hence
\begin{eqnarray}
	&\moy{\xi_u(x, t)} = 0\\
	&\moy{\xi_u(x, t) \xi_u(x', t')} = 2 D \, \delta(t-t') \, \partial_x \, \partial_{x'} \sum_{j=1}^{N} \delta r_j^2 \, \delta(x - r_j) \, \delta(x-x')\label{eqn:varmicropart}
\end{eqnarray}

Let us now compare this noise with the noise $\xi_{u}^D=
\frac{u}{2\sqrt{\rho}} ~\eta$ obtained by linearizing the fluctuating
hydrodynamics in equation~\eqref{eqn:ulinD}. The latter satisfies
\begin{eqnarray}
  \moy{\xi_u^D(x,t) \, \xi_u^D(x',t')}&= \frac{D\, u(x, t) \, u(x', t')}{2\sqrt{\rho(x, t) \rho(x' ,t')}} \,\delta(x-x')\, \delta(t-t')\nonumber\\
  &=  \frac{D\,\partial_x \partial_{x'} \sum_{i,j=1}^{N} \delta r_i \, \delta r_j \, \delta(x-r_i) \, \delta(x'-r_j)}{2 \sqrt{\sum_{k=1}^{N} \delta(x-r_k) \sum_{n=1}^{N} \delta(x'-r_n)}} \delta(x-x') \delta(t-t')\nonumber
\end{eqnarray}
At first glance, this looks different from~\eref{eqn:varmicropart}. As
shown in~\ref{sec:bruit_u_bruit_particules}, these two variances are
however equivalent. This can be seen by noting that the noise
$\xi^D$ appearing in the fluctuating hydrodynamics is
equivalent to $\xi$, i.e.,
\begin{equation}
  \sqrt{\rho(x,t)} \,\eta(x,t) =  \sum_{i=1}^{N} \delta(x-r_i(t)) \, \eta_i(t).
\end{equation}
Linearizing this equality with respect to $r_i$ then gives
\begin{equation}
  \xi_u^D(x,t) = \frac{u(x,t)}{2\sqrt{\rho(x,t)}} ~\eta(x,t) = \sum_{i=1}^{N} \delta r_i \, \partial_x \delta(x-r_i(t)) \, \eta_i(t) = \xi_u(x,t).
\end{equation}

At this stage, we have thus shown that linearizing the fluctuating
hydrodynamics or coarse-graining the microscopic tangent dynamics
yields equivalent equation of evolution for the tangent field $u(x,t)$ for the particle-based noise.

\subsubsection{Noise on the environment}
Let us start by establishing the evolution of the $\delta r_j$,
defined in equation~\eqref{eqn:defdeltar}, linearizing the microscopic
equation~\eqref{eq:bruit sites}:
\begin{eqnarray}
	\td{\delta r_j}{t} 	&= - \int \dd y \, \chi(y, t) \, \delta r_j(t) \, \partial_y \delta(y - r_j(t)) \\
						&= \delta r_j(t) \int \dd y \, \partial_y \xi(y) \, \delta(y - r_j(t))\\
						&= \delta r_j(t) \,\chi'(r_j(t)).
\end{eqnarray}
It\=o lemma applied to the tangent field $u(x,t)=-\sum_{j=1}^N \delta
r_j \partial_x \delta(x-r_j(t))$ defined in equation~\eqref{eqn:defu} then yields:
\begin{eqnarray}\label{bigeq}
  \pd{u}{t}& = \sum_{j=1}^{N} \pd{u}{r_j} \, \chi(r_j) + \sum_{j=1}^{N} \pd{u}{\delta r_j} \, \delta r_j \, \chi'(r_j) + D \sum_{i,j=1}^{N} \frac{\partial^2 u}{\partial r_i \, \partial r_j} \, C(r_i - r_j) \\
	&+ D \sum_{i,j=1}^{N} \frac{\partial^2 u}{\partial r_i \, \partial \delta r_j} \, \delta r_j \, C'(r_j - r_i)
	- D \sum_{i,j=1}^{N} \frac{\partial^2 u}{\partial \delta r_i \, \partial \delta r_j} \, \delta r_i \, \delta r_j \, C''(r_i - r_j)\nonumber
\end{eqnarray}
Let us now show how this equation can be greatly simplified.  First,
since $u$ is linear in the $\delta r_j$'s, we have that
$\partial_{\delta r_j,\delta r_i} u=0$. Furthermore, since
$\partial_{r_i,\delta r_j}u \propto \delta_{i,j}$:
\begin{equation}
  D \sum_{i,j=1}^{N} \frac{\partial^2 u}{\partial r_i \, \partial \delta r_j} \, \delta r_j \, C'(r_j - r_i) \propto C'(0)= 0.
\end{equation}
Moreover,
\begin{eqnarray}
  D \sum_{i,j=1}^{N} \frac{\partial^2 u}{\partial r_i \, \partial r_j} \, C(r_i - r_j) &= -D \sum_{i,j,k=1}^{N} \delta r_k \, \partial_x \partial_{r_i}\partial_{r_j} \delta(x-r_k) \, C(r_i - r_j)\\
  &= -D \sum_{k=1}^{N} \delta r_k \, \partial_x^3 \delta(x-r_k)\, C(0)= D\, \partial_x^2 u.
\end{eqnarray}
Finally, let us consider the two first terms of equation~\eqref{bigeq},
which can be factorized as:
\begin{eqnarray}
  \sum_{j=1}^{N}\Big[ \pd{u}{r_j} \, \chi(r_j) + \pd{u}{\delta r_j} \, \delta r_j \, \chi'(r_j)\Big] = -\sum_{j=1}^N\delta r_j \, \partial_{r_j} [\partial_x \delta(x-r_j)\chi(r_j)].
\end{eqnarray}
We can then use the fact that $\partial_x[\delta
(x-r_j)]\,\chi(r_j)=\partial_x[\delta (x-r_j)\,\chi(x)]$ to get
\begin{eqnarray}
  -\partial_x [\chi(x)\sum_{j=1}^N \delta r_j \,\partial_{r_j} \delta(x-r_j)].
\end{eqnarray}
Finally, we use again that $\partial_{r_j}\delta(x-r_j)=-\partial_x
\delta(x-r_j)$ to obtain the simple form
\begin{equation}
  \partial_x [\chi(x)\sum_{j=1}^N\delta r_j \partial_{x} \delta(x-r_j)]=
  -\partial_x [u(x) \chi(x) ].
\end{equation}
All in all, equation~\eqref{bigeq} simplifies into
\begin{equation}\label{eqn:toto18}
	\partial_{t} u(x,t) = D \, \partial_x^2 u(x,t) - \partial_x \Gamma_u(x,t)\;\;$with$\;\;\Gamma_u(x,t)=u(x, t) \chi(x, t)
\end{equation}
which is exactly the same equation as~\eqref{eqn:FHUNE}, obtained by
linearizing the fluctuating hydrodynamics~\eqref{eq:Dean noise
  environment}. In particular, the noises $\Gamma_u$ and $\Gamma_u^D$
are identical.

\subsubsection{Comparison between particle-based and environment-based tangent dynamics}
Again, we want to show that when the correlation length of the
environment $\ell$ is much shorter than the interparticle distance
$d$, particle-based and environment-based become similar. To do so, we
thus look at the variance of $\Gamma_u(x,t)$ as $d\gg \ell$. 

Starting from equation~\eqref{eqn:toto18} and using the explicit expression
of $u(x,t)$, one gets
\begin{eqnarray}
  \frac{\moy{\Gamma_u(x, t) \, \Gamma_u(x', t')}}{2D}
  &=  \delta(t - t') \,  u(x, t) \, u(x', t') \, C(x-x') \nonumber\\
  &= \delta(t - t') \, \sum_{i,j=1}^{N} \delta r_i \, \delta r_j \, \partial_{x}[\delta(x - r_i)] \, \partial_{x'}[\delta(x' - r_j)] \, C(x-x'). \nonumber
\end{eqnarray}
Using that $\partial_{x}[\delta(x - r_i)]=-\partial_{r_i}[\delta(x -
r_i)]$, this becomes
\begin{eqnarray}
  \frac{\moy{\Gamma_u(x, t) \, \Gamma_u(x', t')}}{2D}              &= \delta(t - t') \, \sum_{i,j=1}^{N} \delta r_i \, \delta r_j \, \partial_{r_{i}}[\delta(x - r_i)] \, \partial_{r_{j}}[\delta(x' - r_j)] \, C(x-x') \nonumber\\
		&= \delta(t - t') \,  \sum_{i,j=1}^{N} \delta r_i \, \delta r_j \, \partial_{r_{i}} \partial_{r_{j}} \Big[\delta(x - r_i) \, \delta(x' - r_j) \, C(r_i-r_j) \Big].\nonumber
\end{eqnarray}
If $d \gg \ell$, $C(r_{i}(t) - r_{j}(t)) = \delta_{i,j} \, C(0) =
\delta_{i,j}$. Then
\begin{eqnarray}
	\moy{\Gamma_u(x, t) \, \Gamma_u(x', t')} 
		&= 2D \, \delta(t - t') \, \Big(\sum_{i,j=1}^{N} \delta r_i \, \delta r_j \, \partial_{r_{i}} \partial_{r_{j}} \Big[\delta(x - r_i) \, \delta(x' - r_j) \, \delta_{i,j} \Big]\Big)\nonumber\\
		&= 2D \, \delta(t - t') \, \partial_x \partial_{x'} \Big[\sum_{i,j=1}^{N} \delta r_i \, \delta r_j \, \delta(x - r_i) \, \delta(x' - r_j) \, \delta_{i,j} \Big]\nonumber\\
		&= 2D \, \delta(t - t') \, \partial_x \partial_{x'} \Big[\sum_{j=1}^{N} \delta r_j^2 \, \delta(x - r_j) \, \delta(x - x') \Big]
\end{eqnarray}
which is the same as the variance $\moy{\xi_u(x, t) \, \xi_u(x', t')}$
of the noise $\xi_u(x,t)$ appearing in the fluctuating ``tangent''
hydrodynamics~\eqref{eqn:toto12} stemming from a particle-based noise.

Note that one difference remains between these two cases: for the
particle-based noise, the $\delta r_j$'s are constant whereas they
evolve for the environment-based noise. Let us now focus on this
latter case to understand the underlying physics. The dynamics on
$\delta r_j$'s read
\begin{equation}
  \delta\dot r_j= \chi'(r_j) \,\delta r_j = \delta r_j  \int \dd y \,\partial_y \delta(y-r_j) \,\chi(y) \equiv  \delta\chi_j. 
\end{equation}
This is a Langevin equation with a multiplicative noise
$\delta\chi_j$. One trivially has $\moy{\delta\chi_j}=0$ while its
correlations are given by
\begin{equation*}
  \moy{\delta\chi_j(t)\,\delta\chi_j(t')} = 2 D \,\delta r_j^2 \,\delta(t-t') \int \dd y \,\dd z \,\partial_y[\delta(y-r_j)] \,\partial_z [\delta (z-r_j)] \,C(y-z).
\end{equation*}
Integrating by part on $y$ and $z$ then yields
\begin{eqnarray}
  \moy{\delta\chi_j(t)\,\delta\chi_j(t')} &= -2 D \,\delta r_j^2 \,\delta(t-t') \int \dd y\, \dd z\, \delta(y-r_j)\,\delta (z-r_j)\, C''(y-z)\nonumber\\
  &= -2 D \,\delta r_j^2 \,\delta(t-t') \,C''(0) = 2 D \,\Big(\frac{\delta r_j}{\ell}\Big)^2\,\delta(t-t').
\end{eqnarray}

If one studies a linearized dynamics at a scale much shorter than the
environment correlation length ($\delta r_j\ll \ell$), the second
cumulant of $\delta\chi_j$ vanishes. The noise on $\delta r_j$ is
identically zero and $\delta \dot r_j=0$. Physically this is
consistent with the fact that two nearby particles separating by a
distance much shorter than $\ell$ experience the same
noise. Conversely, if one studies a tangent dynamics at a scale larger
than $\ell$, the two initial conditions for particle $j$ leads to two
different noise realizations. Particle-based and environment-based
noise then yield the same form of tangent fluctuating hydrodynamics,
but the underlying $\delta r_j$'s have different dynamics.

\subsubsection{From tangent dynamics to Lyapunov exponent}

In this section we have shown that linearizing the microscopic
dynamics and then coarse-graining the corresponding tangent dynamics
is equivalent to starting from the coarse-grained dynamics and
linearizing it. Again, the equivalence between particle-based and
environment-based noises when the correlation length of the
environment is much shorter than the interparticle distance is valid.
The legitimates the approach to compute the large deviation of the
largest Lyapunov exponent in large driven diffusive systems followed
in~\cite{Laffargue2015} which started directly from the fluctuating
hydrodynamics. 

To compute the Lyapunov exponent starting from the tangent dynamics,
it is often convenient to introduced the normalized tangent field $v
\equiv u/\norm{u}$. For particle-based noise, the dynamics of $v$
reads
\begin{eqnarray}
  \partial_t v(x,t) =  &D \,\partial_{x}^2 v(x, t) - \partial_x \Big[\frac{v(x, t)}{2\sqrt{\rho(x,t)}} \eta(x, t) \Big] \\
  &- v(x, t) \int \dd y \, \Big( v(y, t) D \,\partial_{y}^2 v(y, t) - v(y, t) \partial_y \Big[\frac{v(y, t)}{2\sqrt{\rho(y,t)}} \eta(y, t) \Big] \Big)\nonumber
\end{eqnarray}
whereas for the environment-based noise, it reads
\begin{eqnarray}
 \partial_t v(x,t) =  &D \,\partial_{x}^2 v(x, t) - \partial_x \left[v(x, t) \chi(x, t) \right] \\
 &- v(x, t) \int \dd y \, \left( v(y, t) D \,\partial_{y}^2 v(y, t) - v(y, t) \partial_y \left[v(y, t) \chi(y, t) \right] \right).\nonumber
\end{eqnarray}
\revision{Note that we have used standard differential calculus, and not It\=o
stochastic calculus, to derive these formulae. This is legitimate, as
surprisingly as this may seem, since Dean's equation and its
linearized version are identical in both prescriptions. This has often
been asserted in the literature~\cite{kim2014equilibrium} and a
detailed proof can be found in Appendix C of~\cite{solon2015active}.}

The largest Lyapunov exponent is then given by
\begin{equation*}
  \lambda_{part}(t) = \frac{1}{t} \int_{0}^{t} \dd t' \int \dd x \, \Big( v(x, t')\, D \,\partial_{x}^2 v(x, t') - v(x, t) \, \partial_x \Big[\frac{v(x, t')}{2\sqrt{\rho(x,t')}} \eta(x, t') \Big] \Big)
\end{equation*}
and
\begin{equation*}
  \lambda_{env}(t) = \frac{1}{t} \int_{0}^{t} \dd t' \int \dd x \, \Big( v(x, t')\, D \,\partial_{x}^2 v(x, t') - v(x, t) \, \partial_x \Big[v(x, t') \chi(x, t') \Big] \Big)
\end{equation*}
The cumulant generating function corresponding to these Lyapunov
exponents have been computed up to fifth order
in~\cite{Laffargue2015}. Interestingly, the particle-based noise
corresponds to free particles, as expected, while the the
Kipnis-Marchioro-Presutti~\cite{Kipnis1982} model corresponds to a
realization of the environment-based noise.

\section{Lyapunov exponent of the Dean-Kawasaki equation}
\label{sec:interactions}
\subsection{The largest Lyapunov exponent}
Let us now discuss how the approach presented above could be used to
compute the fluctuations of Lyapunov exponents for interacting
systems. We consider the fluctuating hydrodynamics of $N$ particles
interacting via a pair-potential $V$, as derived by
Dean~\cite{Dean1996}:
\begin{eqnarray}
  \pd{\rho({\mathbf x},t)}{t} &= D \,\bnabla_{\mathbf x}^{2} \rho({\mathbf x},t) + \bnabla_{\mathbf x}(\sqrt{\rho({\mathbf x},t)} \, \boldsymbol{\eta}({\mathbf x}x,t))\\
  &\quad + \bnabla_{\mathbf x} \Big[  \int \dd y \, \rho({\mathbf x},t) \,  \bnabla_{\mathbf x}V({\mathbf x}-{\mathbf y}) \rho({\mathbf y}, t) \Big]
\end{eqnarray}
where $\boldsymbol{\eta}({\mathbf x},t)$ is a Gaussian white noise of variance $2D$. As shown
above, this corresponds to particle-based noise as much as to an
environment-based noise in the proper limit. The tangent vector $u$
thus evolves according to
\begin{eqnarray}
  \partial_t u({\mathbf x},t) = &D \, \bnabla_{\mathbf x}^2 u({\mathbf x},t) + \bnabla_{\mathbf x} \Big( \frac{u({\mathbf x},t)}{2 \sqrt{\rho({\mathbf x},t)}} \,\boldsymbol{\eta}({\mathbf x},t) \Big)  \\ &+ \bnabla_{\mathbf x}\Big(\, \int \dd y \,[u({\mathbf x},t)  \, \bnabla_{\mathbf x}V({\mathbf x}-{\mathbf y}) \rho({\mathbf y}, t) +  \rho({\mathbf x},t) \, \bnabla_{\mathbf x}V({\mathbf x}-{\mathbf y})\,u({\mathbf y},t)]\Big)\nonumber
\end{eqnarray}
Again, we introduce a normalized tangent vector $v=u/\norm{u}$, whose
dynamics is given by
\begin{eqnarray}
  \partial_t v({\mathbf x}, t) = & D \,\bnabla_{\mathbf x}^2 v({\mathbf x}, t) + \bnabla_{\mathbf x} \Big( \frac{v({\mathbf x}, t)}{2 \sqrt{\rho({\mathbf x}, t)}} \, \boldsymbol{\eta}({\mathbf x}, t) \Big)\nonumber  \\
  &+ \bnabla_{\mathbf x} \Big(\int \dd y \,[v({\mathbf x}, t) \,\bnabla_{\mathbf x}V({\mathbf x}-{\mathbf y})\rho({\mathbf y}, t)+ \rho({\mathbf x}, t) \,\bnabla_{\mathbf x}V({\mathbf x}-{\mathbf y}) v({\mathbf y},t)]\Big)\nonumber\\
  &- v({\mathbf x}, t) \int \dd z \Bigg\{ D \, v({\mathbf z}, t) \, \bnabla_{\mathbf z}^2v({\mathbf z}, t) + v({\mathbf z}, t) \, \bnabla_{\mathbf z}\Big[ \frac{v({\mathbf z}, t)}{2 \sqrt{\rho({\mathbf z}, t)}} \,\boldsymbol{\eta}({\mathbf z}, t) \Big] \nonumber \\
  &+ v({\mathbf z}, t) \, \bnabla_{\mathbf z} \Big(\int \dd y \,[v({\mathbf z}, t) \,\bnabla_{\mathbf z}V({\mathbf z}-{\mathbf y})\rho({\mathbf y}, t)+\rho({\mathbf z}, t) \,\bnabla_{\mathbf z}V({\mathbf z}-{\mathbf y})\,v({\mathbf y}, t)]\Big)\Bigg\}\nonumber
\end{eqnarray}
meaning that the largest Lyapunov exponent is given by
\begin{eqnarray}
  \lambda(t) = &\frac{1}{t} \int_{0}^{t} \dd t' \int \dd x \Bigg\{ D\, v({\mathbf x}, t') \,\bnabla_{\mathbf x}^2 v({\mathbf x}, t') + v({\mathbf x}, t') \, \bnabla_{\mathbf x}\Big[ \frac{v({\mathbf x}, t')}{2 \sqrt{\rho({\mathbf x}, t')}} \boldsymbol{\eta}({\mathbf x}, t') \Big]\nonumber  \\
  &+ v({\mathbf x}, t') \,\bnabla_{\mathbf x}\Big(\int \dd y [v({\mathbf x}, t') \,\bnabla_{\mathbf x}V({\mathbf x}-{\mathbf y})\,\rho({\mathbf y}, t')+ \rho({\mathbf x}, t') \, \bnabla_{\mathbf x}V({\mathbf x}-{\mathbf y}) v({\mathbf y}, t')]\Big)\Bigg\}\nonumber
\end{eqnarray}
This explicit formula for the largest Lyapunov exponent could then be
used to compute its cumulant-generating function, following the path
set in~\cite{Laffargue2015}.

\subsection{Mean field}
The simplest approximation scheme that can be implemented consists in
retaining Gaussian fluctuations for the density field $\rho$ and the
tangent vector $v$. This section is devoted to establishing the
feasibility of such an approximation. Motivations can be found in the
statics of simple classical or quantum fluids, where this is called
the Random Phase Approximation (RPA), and we shall keep this name in
what follows. A recent work in which this approximation has proved
useful within a dynamic framework can be found here \cite{Dean2014} or
\cite{Demery2014} (this includes a discussion on the range of validity
of the RPA approximation, which becomes exact in specific limiting
cases). Here we linearize the dynamics of both the $\rho$ and $v$
fields, and use this as the simplest possible approximation. It would
certainly be interesting to understand if this corresponds to a
limiting case for the tangent dynamics as well. The expansion is
carried out in powers of $\psi = \rho - \rho_0$ and $\chi = v - v_0$,
where $v_0(\mathbf{x})$ is the normalized tangent field obtained
within a straight mean-field approximation. We begin with the
evolution equation for the normalized tangent vector $v$, which reads,
with condensed notations,
\begin{equation}
\partial_t v = A v - v \int v.Av
\end{equation}
where $A$ is a linear operator acting upon $v$. It is important to note that $A$ functionally depends on the density field $\rho$ and on the external noise $\boldsymbol{\xi}$. We find that
\begin{eqnarray}\label{eq:evol-v-DK}
Av(\mathbf{x},t)=&\boldsymbol{\nabla} \cdot \left[D \boldsymbol{\nabla} v(\mathbf{x},t)+\rho(\mathbf{x},t)\int_{\mathbf{y}} \boldsymbol{\nabla} V(\mathbf{x} - \mathbf{y})v(\mathbf{y}, t)\right.\nonumber
\\
&\left.+v(\mathbf{x},t)\int_{\mathbf{y}}\bnabla V(\mathbf{x}-\mathbf{y}) \rho(\mathbf{y},t)+\sqrt{2D}\frac{v(\mathbf{x},t)}{2\sqrt{\rho}}\boldsymbol{\xi}(\mathbf{x},t)\right]
\end{eqnarray}
where $\boldsymbol{\xi}=(2D)^{-1/2}\boldsymbol{\eta}$ is a delta
correlated Gaussian white noise. Within a mean-field approximation,
the density field $\rho$ is replaced with its uniform average
$\rho_0$, and the noise is neglected. With these simplifications
\eref{eq:evol-v-DK} leads, in Fourier space, to
\begin{equation}\label{eq:evol-v-mf}
\partial_t v(\mathbf{k},t)=-\Omega_{\mathbf{k}} v(\mathbf{k},t)+v(\mathbf{k},t)\int_{\mathbf{k}'}v(-\mathbf{k}',t)\Omega_{\mathbf{k}'} v(\mathbf{k}',t)
\end{equation}
where $\Omega_{\mathbf{k}}=D k^2(1+\beta\rho_0 V(\mathbf{k}))$
($\beta=D^{-1}$ is an inverse temperature), and where $V(\mathbf{k})$
is the Fourier transform of the interaction potential, namely
$V({\mathbf k})=\int_{L^3}\dd^3
x\ee^{-i\mathbf{k}\cdot\mathbf{x}}V(\mathbf{x})$. We assume the system
to be enclosed in a cubic box of linear size $L$ with periodic
boundary conditions, hence the inverse Fourier transform is given by
$V(\mathbf{x})=\int_{\mathbf k}
\ee^{i\mathbf{k}\cdot\mathbf{x}}V(\mathbf{k})$, with
$\int_\mathbf{k}=L^{-3}\sum_{\mathbf{k}}$, and the $\mathbf k$'s
components are integer multiples of $2\pi/L$. The largest Lyapunov
exponent itself is given in terms of the stationary solution
$v_0(\mathbf{k})$ of \eqref{eq:evol-v-mf} by
\begin{equation}
\lambda_M=-\int_{\mathbf{k}'}v_0(-\mathbf{k}')\Omega_{\mathbf{k}'} v_0(\mathbf{k}')
\end{equation}
Denoting by $\mathbf{q}$ the wave vector that minimizes $\Omega_\mathbf{k}$, it is immediate to realize that
\begin{equation}
v_0(\mathbf{x})=W\cos\mathbf{q}\cdot\mathbf{x},\; v_0(\mathbf{k})=W^{-1}\left(\delta_{\mathbf{k},-\mathbf{q}}+\delta_{\mathbf{k},\mathbf{q}}\right)
\end{equation}
where $W=\sqrt{2/L^3}$ is a normalizing factor that ensures $\int_\mathbf{x} v_0^2(\mathbf{x})=1$. Since we have in mind a smooth potential, like that of interacting harmonic spheres that has been recently used in many studies of glass-formers, and which has $V(r)=\eps\theta(\sigma-r)(1-r/\sigma)^2$, and $V(\mathbf{k})=8\pi\eps\sigma^3\frac{2 k\sigma+k\sigma\cos k\sigma-3\sin k\sigma}{(k\sigma)^5}$, it is clear that $\mathbf{q}=\pm\frac{2\pi}{L}\mathbf{e}$, where $\mathbf{e}$  is either of three basis unit vectors. The resulting largest Lyapunov exponent then reads, in the large system size limit,
\begin{equation}
\lambda_M=-\frac{4\pi^2}{L^2}D\left[1+\beta\rho_0 \frac{2 \pi  \eps \sigma ^3}{15}\right]
\end{equation}
We have used this harmonic sphere interaction for concreteness, but our result for $\lambda=-\Omega_\mathbf{q}$ is of course more general.

With the reference fields around which to expand now at hand, namely
$\rho_0$ and $v_0$, here is how we could set up a Gaussian (RPA) expansion around mean-field. We would start by writing the evolution equation for $\chi = v -
v_0$ to linear order in $\chi$, $\psi = \rho - \rho_0$, and in the
noise $\boldsymbol{\xi}$. We now address how the average Lyapunov and its fluctuations get renormalized \revision{by quadratic fluctuations around mean-field.

\subsection{Gaussian fluctuations}
The linearized dynamics for $\chi=v-v_0$ is given by
\begin{eqnarray}
\partial_t \chi = &A_0 \chi - \chi \int v_0 A_0 v_0 - v_0 \int \left( v_0 A_0 \chi + \chi A_0 v_0 \right)\nonumber\\
&+\delta A v_0 - v_0 \int v_0 \delta Av_0
\end{eqnarray}
where $A_0$ is the operator $A$ evaluated at zero noise and at
$\psi=\rho-\rho_0 = 0$, while $\delta A$ is the linear correction to
$A$ in an expansion in the $\psi$ and $\boldsymbol{\xi}$ fields. The
action of $A_0$ and $\delta A$ on an arbitrary field $f({\mathbf x})$
with Fourier transform $f(\mathbf{k})$ is explicitly given by
\begin{eqnarray}
A_0f(\mathbf{k})&=-\Omega_{\mathbf{k}} f(\mathbf{k})\\
\delta A f(\mathbf{k})&= i \mathbf{k} \cdot \int_{\mathbf{k}'} \Big[ i \mathbf{k}' V(\mathbf{k}') (\psi(\mathbf{k}-\mathbf{k}',t) f(\mathbf{k}')\nonumber\\
&\qquad + f(\mathbf{k}-\mathbf{k}')\psi(\mathbf{k}'))+\frac{\sqrt{2D}}{2\sqrt{\rho_0}} f(\mathbf{k}')\boldsymbol{\xi}(\mathbf{k}-\mathbf{k}',t)\Big]
\end{eqnarray}

Using the explicit form of $v_0$, we thus find that the Fourier modes
of $\chi$ evolve according to
\begin{eqnarray}
\partial_t\chi(\mathbf{k},t) = &(\Omega_\mathbf{q}-\Omega_\mathbf{k}) \chi(\mathbf{k},t) +\delta A v_0 \nonumber\\
& + 2v_0(\mathbf{k})\int_{\mathbf{k}'} v_0(-\mathbf{k}')\Omega_{\mathbf{k}'}\chi(\mathbf{k}',t) - v_0 \int v_0 \delta Av_0\label{eqn:D0}
\end{eqnarray}
The symbol $\int_{\mathbf{k}'}$ actually stands for
$L^{-3}\sum_{\mathbf{k}'}$.  For $\mathbf{k}\neq\pm\mathbf{q}$, the
second line of eq.~\eqref{eqn:D0} vanishes and the
evolution equation for $\chi$ can be integrated, to yield
\begin{eqnarray}
\chi(\mathbf{k},t)=\int_{-\infty}^t \dd\tau\ee^{-(\Omega_\mathbf{k}-\Omega_\mathbf{q})(t-\tau)}\delta A v_0(\mathbf{k},\tau)
\end{eqnarray}
which converts into the following time-Fourier expression
\begin{eqnarray}\label{expr-chi-1}
\chi({\mathbf k},\omega)=\frac{\delta A v_0({\mathbf k},\omega)}{-i\omega +\Omega_{{\mathbf k}}-\Omega_{\mathbf q}}
\end{eqnarray}
where
\begin{eqnarray}\label{expr-chi-2}
\delta A v_0({\mathbf k},\omega)=L^{-3}W^{-1}\sum_{{\mathbf p}=\pm{\mathbf q}}\left[\alpha_{{\mathbf k},{\mathbf p}}\psi({\mathbf k}-{\mathbf p},\omega)+\frac{\sqrt{2D}}{2\sqrt{\rho_0}}i{\mathbf k}\cdot\boldsymbol{\xi}({\mathbf k}-{\mathbf p},\omega)\right]
\end{eqnarray}
and the symbol  $\alpha_{{\mathbf k},{\mathbf p}}=i{\mathbf k}\cdot[i({\mathbf k}-{\mathbf p}) V({\mathbf k}-{\mathbf p})+i{\mathbf p} V({\mathbf p})]$. We will be using the statistical properties of $\psi$, namely that
\begin{eqnarray}\label{eqn:truc2}
\langle\psi({\mathbf k},\omega)\psi({\mathbf k}',\omega')\rangle=L^3\delta_{{\mathbf k}+{\mathbf k}',\mathbf{0}}\frac{2D\rho_0{\mathbf k}^2}{\omega^2+\Omega_{\mathbf k}^2}2\pi \delta(\omega+\omega'),
\end{eqnarray}
and
\begin{eqnarray}\label{eqn:truc1}
  \langle\psi({\mathbf k},\omega){\boldsymbol{\xi}}({\mathbf k}',\omega')\rangle= L^3\delta_{{\mathbf k}+{\mathbf k}',\mathbf{0}}\frac{\sqrt{2D\rho_0}i\mathbf{k}}{-i\omega +\Omega_{\mathbf k}} 2\pi \delta(\omega+\omega')
\end{eqnarray}
to determine those of $\chi$ as given by \eqref{expr-chi-1} and
\eqref{expr-chi-2}. Equations~\eqref{eqn:truc2}-\eqref{eqn:truc1} stems from the
dynamics of $\psi$
\begin{equation}
  \dot \psi({\bf k},t) = - \Omega_{\bf k} \psi({\bf k},t) + \sqrt{2 D \rho_0 } i {\bf k} \cdot {\boldsymbol \xi}
\end{equation}
The ${\mathbf k}=\pm{\mathbf q}$ modes of $\chi$ can be seen to
satisfy the following evolution equation,
\begin{eqnarray}
\partial_t[\chi({\mathbf q},t)+\chi(-{\mathbf q},t)]=\Omega_{\mathbf q}(\chi({\mathbf q},t)+\chi(-{\mathbf q},t))
\end{eqnarray}
which tells us that $\chi(\pm{\mathbf q},t)$ is purely imaginary. This is of course consistent with the constraint $\int v^2=1$ when the latter is expressed to linear order in $\chi$.\\

The expression of the leading fluctuating correction to the largest
Lyapunov exponent is given by $\delta\lambda=\delta\lambda$, with
\begin{eqnarray}\label{eq:dlambda}
\delta\lambda=\frac{1}{t}\int_0^t \dd t\int_{\mathbf x}\left[\chi A_0 v_0+v_0 A_0\chi+v_0\delta A v_0\right]
\end{eqnarray}
As will be shown, $\delta\lambda$ renormalizes the fluctuations of
$\lambda$ but not its mean value. The first two integrals in
eq.~\eqref{eq:dlambda} yield a contribution proportional to
$\chi({\mathbf q},t)+\chi(-{\mathbf q},t)$ and thus vanish. The
remaining integral can be expressed in terms of the $\psi$ and
${\boldsymbol{\xi}}$ fields, as
\begin{eqnarray}
t\delta\lambda=&L^{-3}\int\dd t\Bigg(\frac{\alpha_{-{\mathbf q},{\mathbf q}}}{2}(\psi(2{\mathbf q},t)+\psi(-2{\mathbf q},t))\nonumber\\
&+\frac{\sqrt{2D}}{4\sqrt{\rho_0}}i{\mathbf q}\cdot({\boldsymbol{\xi}}(2{\mathbf q},t)-{\boldsymbol{\xi}}(-2{\mathbf q},t))\Bigg)
\end{eqnarray}
Being linear in the fields, the distribution of $\delta\lambda$ is of
course Gaussian (but it doesn't have much meaning anyhow to talk about
high order cumulants given the Gaussian nature of the RPA). After
tedious but standard manipulation, its variance is found to be
\begin{eqnarray}
  t\langle\delta\lambda^2\rangle=L^{-3}\frac{D{\mathbf q}^2}{4\rho_0}\left[1+\frac{4\alpha_{-{\mathbf q},{\mathbf q}}\rho_0}{\Omega_{2 \mathbf {q}}}\right]^2
\end{eqnarray}
The coefficient $\alpha_{-{\mathbf q},{\mathbf q}}$ has the expression $\alpha_{-{\mathbf q},{\mathbf q}}=q^2(V({\mathbf q})-2 V(2{\mathbf q}))$, so that within the RPA approximation,
\begin{eqnarray}\label{var-deltalambda1}
t\langle\delta\lambda^2\rangle=L^{-3}\frac{D{\mathbf q}^2}{4\rho_0}\left[1+\frac{\beta\rho_0(V({\mathbf q})-2 V(2{\mathbf q}))}{1+\beta\rho_0 V(2{\mathbf q})}\right]^2
\end{eqnarray}
The result appearing in \eqref{var-deltalambda1} is interesting in
various respects. First of all, we have not used any fluctuating
hydrodynamics but our expression extrapolated to noninteracting
particles coincides with the exact result obtained in
\cite{Laffargue2015}. Second, it shows the connection between the
Lyapunov exponent and the microscopic interaction between particles,
with higher multiples of the slowest mode ${\mathbf q}$ entering
higher order fluctuations of $\lambda$. This too, though in a
different fashion, was seen in \cite{Laffargue2015}. Using our pet
harmonic sphere potential, which is mildly repulsive, we find
fluctuations of chaoticity to be reduced with respect to those of an
ideal gas (because $V(\mathbf{q})-2 V(2\mathbf{q})<0$, a property that
other very short range potentials share, such as a screened Coulomb
potential for instance). Again, a similar trend can be found in low
density lattice gases, but there fluctuations shoot up as the density
exceeds a threshold value. It would be dangerous to try and extrapolate
the RPA result into a dense regime.}

\section{Conclusion}

In this article, we have presented two different approaches to compute the
Lyapunov exponent of stochastic systems, going beyond the standard
``same-noise'' vs ``different-noise'' paradigm usually referred
to~\cite{Benzi1985,Arnold1986,Arnold1988}. Indeed, we show that
enforcing the same noise realization in two stochastic processes is
ambiguous, something which is well-known in the damage spreading
community~\cite{Grassberger1995a, Hinrichsen1997} but had not been pointed out in the much
simpler context of particles diffusing in a fluid. Even without taking
into account the full consequences of hydrodynamics, the fact that the
noise on a colloidal particle comes from collisions with fluid
particles requires a new prescription when comparing two initial
conditions experiencing ``the same noise''. 

We have shown that the Lyapunov exponent computed using this
environment-based noise have different distributions than the ones
stemming from the standard particle-based noise. When the correlation
length of the environment is much shorter than the interparticle
distance, however, the two approaches become equivalent (as they should).

When considering collective modes, like the density field, we have
shown that linearizing the fluctuating hydrodynamics is equivalent to
linearizing the microscopic dynamics and then coarse-graining the
tangent dynamics. In section~\ref{sec:interactions}, we have studied
the case of interacting particles, providing a mean-field estimate for
the largest Lyapunov exponent. 

Our article thus both provides a starting point for future studies of
fluctuations of Lyapunov exponents in large interacting stochastic
systems and highlights the underlying hypothesis made on the origin of
noise in stochastic processes and their importance when dealing with
chaos.

\appendix{}

\section{Time discretization for the environment-based noise}
\label{app:NTD}
When time-discretizing the stochastic differential
equation~\eref{eqn:SDE1} during a small time interval $\Delta t$, one
needs to specify at which time, $t+\eps \Delta t \in [t,t+\Delta t]$,
the prefactor of the noise is evaluated:
\begin{eqnarray}
  r(t+\Delta t)&=r(t) + \int_t^{t+\Delta t} \int \dd y \, \delta(y-r(t')) \,\chi(y,t') \,\dd t'\nonumber\\
  &\simeq r(t)+ \int \dd y \,\delta(y-r(t+\eps \Delta t)) \int_t^{t+\Delta t}  \,\chi(y,t') \,\dd t'.\label{eqn:A1}
\end{eqnarray}
As we now show, the Fokker-Planck equation corresponding
to~\eref{eqn:SDE1} is actually independent of $\eps$. Indeed,
\eref{eqn:A1} amounts to
\begin{eqnarray}
  \Delta r \simeq  \int_{t}^{t + \Delta t}\dd t' \int \dd y \, \delta(y - r(t) - \eps \Delta r)) \, \chi(y, \,t')
\end{eqnarray}
which is a self-consistent equation on $\Delta r$. As $\Delta r \ll 1$, we can perform a series expansion of the Dirac distribution in the integral
\begin{equation}
	\Delta r \simeq  \int_{t}^{t + \Delta t}\!\!\!\!\!\!\!\!\!\!\dd t' \int \dd y \, \Big[ \delta(y - r(t)) \, \chi(y, \,t') - \eps \Delta r \, \chi(y, \,t') \, \partial_{y}\delta(y - r(t)) + \cdots \Big]\label{eqn:A2}
\end{equation}
Replacing $\Delta r$ in the integral by its expression~\eref{eqn:A2} then yields, when $\Delta t\to 0$,
\begin{equation}
  \frac{\moy{\Delta r}}{\Delta t} \simeq - \frac{\eps}{\Delta t} \int_{t}^{t + \Delta t}\!\!\!\!\!\!\!\!\!\!\dd t' \int_{t}^{t + \Delta t}\!\!\!\!\!\!\!\!\!\!\dd t'' \int \!\!\dd y \int \!\!\dd y' \, \partial_{y}\delta(y - r(t)) \moy{\chi(y, \,t') \,  \chi(y', \,t'')}.
\end{equation}
Using $\langle \chi(y,t')\,\chi(y',t'')\rangle=C(y-y') \delta(t'-t'')$ and integrating over $t'$ and $t''$ then yields
\begin{eqnarray}      
  \frac{\moy{\Delta r}}{\Delta t}&= 2 D \eps \int \dd y \int \dd y' \, \delta(y - r(t))  \, \delta(y' - r(t))\,C'(y - y') + \underset{\Delta t \to 0}{\mathcal{O}}(\Delta t)\nonumber\\
	&= 2 D \eps \, C'(0) + \underset{\Delta t \to 0}{\mathcal{O}}(\Delta t).
\end{eqnarray}
Finally, when $\Delta t$ tends to $0$, the limit of the moment rate is
\begin{equation}
  \lim_{\Delta t \to 0} \frac{\moy{\Delta r}}{\Delta t} = 2 D \eps \, C'(0).
\end{equation}
Since $C$ is even, $C'(0) = 0$, and the first moment rate, which is
also the first coefficient of the Kramers-Moyal expansion, is zero
for any choice of time-discretization. All the discretization $\eps
\in [0,\,1]$ are then equivalent.

\section{Stratonovitch calculus}
\label{app:strato}
The previous Langevin equation is independent of the discretization.  In this section, we will consider this equation as a Stratonovitch equation.

The local density is defined by
\begin{equation}
	\rho(x, t) = \sum_{j=1}^{N} \delta(x-r_{j}(t)).
\end{equation}
In order to find the Langevin equation the density evolves according with, we perform a Kramers-Moyal expansion of the moments $\Delta \rho = \rho(t + \Delta t) - \rho(t)$, averaged over the realization of the noise during the time interval $[t, t + \Delta t]$, at fixed value of $\rho(t)$. We use the Stratonovitch discretization that allows us to manipulate singular functions within the framework of standard differential calculus, so that
\begin{equation}
	\partial_{t}\rho = - \sum_{j=1}^{N} \int \dd y ~\chi(y, t) ~\delta(y - r_{j}) \partial_{x} \delta(x - r_{j})
\end{equation}
which actually means that:
\begin{equation*}
	\Delta \rho = - \sum_{j=1}^{N} \int \dd y ~\delta(y - r_{j}(t) - \frac{\Delta r_{j}}{2}) ~\partial_{x} \delta(x - r_{j}(t) - \frac{\Delta r_{j}}{2}) \int_{t}^{t+\Delta t} \!\!\!\!\!\!\!\!\!\!\dd \tau ~\chi(y, \tau).
\end{equation*}
Once expanded to leading order in $\Delta t$, given that
\begin{equation}
	\Delta r_{j} = \int \dd y ~\delta(y - r_{j} - \frac{\Delta r_{j}}{2}) \int_{t}^{t+\Delta t} \!\!\!\!\!\!\!\!\!\!\dd \tau ~\chi(y, \tau),
\end{equation}
we arrive at
\begin{equation*}
	\lim_{\Delta t \to 0} \frac{\moy{\Delta \rho}}{\Delta t} = D \sum_{j=1}^{N} \int \dd y ~\dd y'
\left[ \delta(y - r_{j}) ~\partial_{x}^2 \delta(x - r_{j}) + \partial_{y} \delta(y - r_{j}) ~\partial_{x} \delta(x - r_{j}) \right] C(y-y')
\end{equation*}
which leads to
\begin{equation}
	\lim_{\Delta t \to 0} \frac{\moy{\Delta \rho}}{\Delta t} = D ~\partial_{x}^{2} \rho
\end{equation}
since $C(0) = 1$ and $C'(0) = 0$.
Similarly, we find that
\begin{equation*}
	\lim_{\Delta t \to 0} \frac{\moy{ \Delta \rho(x, t) \Delta \rho(x', t')}}{\Delta t} =  \!\!\!\sum_{i,j=1}^{N}  \!\int \!\!\dd y ~\dd y' \delta(y - r_{i}) \delta(y' - r_{j}) \partial_{x} \delta(x - r_{i}) \partial_{x'} \delta(x' - r_{j}) C(y-y')
\end{equation*}
which gives us
\begin{eqnarray}
	\lim_{\Delta t \to 0} \frac{\moy{ \Delta \rho(x, t) \Delta \rho(x', t')}}{\Delta t} &= 2 D\sum_{i,j=1}^{N} \partial_{x} \partial_{x'} \left( \delta(x - r_{i}) \,\delta(x' - r_{j}) \,C(r_{i} - r_{j}) \right)\nonumber\\
	&= 2 D ~\partial_{x} \partial_{x'} \left( \rho(x, t) \,\rho(x', t) \,C(x-x') \right).
\end{eqnarray}
We can now express the Langevin equation (in Ito's discretization) for the density:
\begin{equation}
	\partial_{t} \rho = D ~\partial_{x}^{2} \rho - \partial_x \Gamma(x, t)
\end{equation}
with
\begin{eqnarray}
	\moy{\Gamma(x, t)} &= 0\\
	\moy{\Gamma(x, t) \, \Gamma(x', t')} &= 2 D \, \rho(x, t) \, \rho(x', t) \, C(x-x')
\end{eqnarray}

\section{Tangent dynamics of particle-based noise: hydrodynamics derivation versus linearization}
\label{sec:bruit_u_bruit_particules}
In this appendix we show that the noise
\begin{equation}
	\xi_{u}^{D}(x, t) = \left[ \frac{u(x,t)}{2\sqrt{\rho(x,t)}} ~\eta(x, t) \right]
\end{equation}
obtained by linearizing the fluctuating hydrodynamics
\begin{equation}
  \dot\rho(x,t) = \partial_x^2 \rho(x,t) - \partial_x [\sqrt{\rho(x,t)} \,\eta(x,t)]
\end{equation}
is equivalent to the one obtained by coarse-graining the microscopic
tangent dynamics:
\begin{equation}
	\xi_{u}(x, t) = \partial_{x} \sum_{j=1}^{N} \delta r_j \, \delta(x-r_j(t)) \,\eta_j(t).
\end{equation}
Since these noises are Gaussian, we simply have to show that their
mean and variance are equal. Both mean are trivially zero, and we thus
only consider their variances.

In section \ref{sec:noiseindpart}, we showed that the noise
\begin{equation}
  \xi(x, t,\{r_j\}) = \displaystyle\sum_{j=1}^{N} \rho_j(x, t) \,\eta_j(t)  = \displaystyle\sum_{j=1}^{N} \delta(x-r_j(t)) \,\eta_j(t)
\end{equation}
is equivalent to
\begin{equation}
	\xi^{D}(x, t,\{r_j\}) = \partial_x \Big[ \sqrt{\rho(x, t)} \,\eta(x, t) \Big]=\Big[ \Big(\sum_{j=1}^{N} \delta(x - r_{j}(t)\Big)^{1/2} \,\eta(x, t) \Big]
\end{equation}
where we have explicitly written the dependence of $\xi$ and $\xi^D$
on $r_j$.

We will now linearize $\xi$ and $\xi^D$ with respect to the $r_j$'s and show
that the corresponding linearized noises correspond to $\xi^D_u$ and
$\xi_u$. Since $\xi=\xi^D$, this will establish the equivalence
between $\xi_u$ and $\xi^D_u$

We first look at
\begin{equation*}
  \mathcal{C}(x, x', t, t') \equiv \moy{\Big[\xi(x, t, \{r_j + \delta r_j\}) - \xi(x, t, \{r_j\})\Big]\, \Big[\xi(x', t', \{r_j + \delta r_j\}) - \xi(x', t', \{r_j\})\Big]}
\end{equation*}
Using the explicit expression of $\xi$, one gets
\begin{eqnarray}
	\frac{\mathcal{C}(x, x', t, t')}{2D} &= \sum_{i, j=1}^{N} \Big[ \delta(x - r_i(t) - \delta r_i) - \delta(x - r_i(t)) \Big] \nonumber\\
	&\hskip1cm \times \Big[ \delta(x' - r_j(t') - \delta r_j) - \delta(x' - r_j(t')) \Big] \frac{\moy{\eta_i(t) \eta_{j}(t')}}{2D}\nonumber\\
	&= \sum_{i,j=1}^{N} \Big[ \delta(x - r_i(t) - \delta r_i) - \delta(x - r_i(t)) \Big] \nonumber\\
	&\hskip1cm \times \Big[ \delta(x' - r_j(t) - \delta r_j) - \delta(x' - r_j(t)) \Big] \delta_{i,j}\,\delta(t-t')\nonumber\\
	&= \delta(t-t')\sum_{i=1}^{N} \Big[ \delta(x - r_i - \delta r_i) - \delta(x - r_i) \Big] \, \Big[ \delta(x' - r_i - \delta r_i) - \delta(x' - r_i) \Big]\nonumber\\
	&\simeq \delta(t-t')\sum_{i=1}^{N} \Big[\delta r_i^2 \,\partial_x \delta(x - r_i) \,\partial_{x'} \delta(x' - r_i) + \mathcal{O}(\delta r_i^3)\Big].
\end{eqnarray}
The leading order in $\delta r$ of $\mathcal{C}(x, x', t, t')$ is exactly the correlation of $\xi_u(x, t)$. 

Let us now calculate the same quantity for $\xi^{D}(x,t)$:
\begin{eqnarray}
	\frac{\mathcal{C}^D(x, x', t, t')}{2D} &= \Bigg(\Big[\sum_{i=1}^{N} \delta(x - r_i(t) - \delta r_i)\Big]^{1/2} - \Big[\sum_{i=1}^{N}\delta(x - r_i(t))\Big]^{1/2} \Bigg) \nonumber\\
	&\hskip1cm \times \Bigg(\Big[\sum_{i=1}^{N} \delta(x' - r_i(t') - \delta r_i)\Big]^{1/2} - \Big[\sum_{i=1}^{N}\delta(x' - r_i(t'))\Big]^{1/2} \Bigg) \nonumber\\
	&\hskip1cm \times \frac{\moy{\eta(x, t) \,\eta_(x', t')}}{2D}\nonumber\\
	&= \delta(x - x')\delta(t-t') \, \Bigg[ \underbrace{\sum_{i=1}^{N} \delta(x - r_i - \delta r_i)}_{\fbox{1}} + \textcolor{blue}{\,\sum_{i=1}^{N} \delta(x - r_i)} \nonumber\\
	&\hskip3cm - \underbrace{2 \Big( \sum_{i,j=1}^{N} \delta(x - r_i - \delta r_i) \delta(x - r_i) \Big)^{1/2}}_{\fbox{3}}\Bigg].\nonumber
\end{eqnarray}
We now perform a series expansion to second order in $\delta r_j$. The
term \fbox{1} gives
\begin{equation}
	\sum_{i=1}^{N} \Big[ \color{blue}{\delta(x - r_i)} \textcolor{red}{\,- \,\delta r_i \,\partial_{x} \delta(x - r_i)} \textcolor{DarkGreen}{\,+ \,\frac{1}{2} \,\delta r_{i}^{2} \, \partial_{x}^2 \delta(x - r_i)\Big]}
\end{equation}
whereas the term \fbox{3} yields
\begin{equation*}
	\textcolor{blue}{2 \rho}  \textcolor{red}{\,- \sum_{i=1}^{N} \delta r_i \, \partial_x \delta(x - r_i)} \textcolor{DarkGreen}{\,+ \,\frac{1}{2} \sum_{i=1}^{N} \delta r_{i}^2 \, \partial_x^2 \delta(x - r_i)} + \frac{1}{4\rho} \Big[ \sum_{i=1}^{N} \delta r_i \, \partial_x \delta(x - r_i) \Big]^2.
\end{equation*}
Noting that $\rho = \sum_{i=1}^{N} \delta(x - r_i)$, the terms of the
same color cancel and one gets
\begin{eqnarray}
	\frac{\mathcal{C}^D(x, x', t, t')}{2D} & \simeq \frac{1}{4\rho(x,t)} \Big[ \sum_{i=1}^{N} \delta r_i \, \partial_x \delta(x - r_i) \Big]^2 \, \delta(x-x') \, \delta(t - t') + \mathcal{O}(\delta r^3)\nonumber\\
	 									   & \simeq \frac{u(x, t)^2}{4 \rho(x, t)} \, \delta(x-x') \, \delta(t - t') + \mathcal{O}(\delta r^3)\nonumber\\
	 									   & \simeq \frac{u(x,t)}{2\sqrt{\rho(x, t)}} \frac{u(x',t')}{2\sqrt{\rho(x', t')}}\, \delta(x-x') \, \delta(t - t') + \mathcal{O}(\delta r^3)
\end{eqnarray}
which is, at the leading order in $\delta r$, the correlation of
$\xi_u^D(x, t)$. Since $\xi(x, t) = \xi^{D}(x, t)$, we have
$\mathcal{C}(x, x', t, t') = \mathcal{C}^D(x, x', t, t')$, which
implies that $\xi_u(x, t)$. and $\xi_u^D(x, t)$ have the same
correlations, and are thus equivalent.

\section*{References}
\bibliography{Flo}
\end{document}